\documentclass[12pt]{article}

\usepackage{graphicx}

\usepackage[fleqn]{amsmath}
\usepackage{amssymb,latexsym,multirow}
\usepackage{slashbox}
\usepackage{psfrag}

\begin{document}

\title{Approximate self-similar solutions for the boundary value problem arising in modeling of gas flow in porous media with pressure dependent permeability}

\author{Nikolay A. Kudryashov, Mark B. Kochanov, Viktoria L. Savatorova}

\date{Department of Applied Mathematics, National Research Nuclear University
MEPHI, 31 Kashirskoe Shosse, 115409 Moscow, Russian Federation}

\maketitle

\begin{abstract}
The flow of a gas through porous medium is considered in the case of pressure dependent permeability.
Approximate self-similar solutions of the boundary-value problems are found.

\end{abstract}

\section{Introduction}

The investigation of gaseous flows through tight porous media has drawn huge attention because of its practical importance when dealing with extraction of hydrocarbon gases from unconventional gas reservoirs, such as shale gas and coal bed methane reservoirs.
The quantitative analysis of the gas flow may be needed for environmental protection, for example in the areas of contaminant transport and remediation in the unsaturated zone.
There is considerable number of results, indicating that the the range of applicability of conventional Darcy's law has many limitations due to the nature of the porous medium and peculiarities of the flow regimes \cite{Raj2007}.That is why investigation of deviations from Darcy's law becomes a topic of special interest in many studies and applications.There can be distinguished several cases, where the equations, describing gas flow through the porous medium, become non-linear and deviate from the Darcy's law.

Firstly, let us look at the violation of Darcy's law in the case, where gas velocity is relatively high.
This class of flows is discussed in publications \cite{Pascal_1988, Pascal_1991, Pascal_1997, Kudryashov_1980, Kudryashov_1982, Kudryashov_1985}.
It has been shown there that, where Reynolds number for gas is big enough, linear correlation between the pressure gradient and velocity is no longer valid, and the nonlinear inertial term in the momentum equation can no longer be neglected.
Since we are interested in gas filtration, the nonlinear inertial effects must be considered along with the compressibility one.

Thus one-dimentional problem for the isothermal gas flow through a porous medium under the conditions of turbulent filtration was studied in \cite{Kudryashov_1980}.
The solution was obtained in terms of cylindrical functions.
In \cite{Kudryashov_1982}, these results were extended to the case of axisymmetric motion.
Analytical solutions for one-dimensional problems of gas flow for quadratic resistance law were found in \cite{Kudryashov_1985}.

In paper \cite{Pascal_1988}, the conditions for existence of self-similar solutions for the momentum equations, describing unsteady flow through a porous medium and incorporating the nonlinear inertial term, are examined.
According to \cite{Pascal_1988}, self-similar solutions for non-Darcy gas flow exist, if we assume a constant pressure or a constant mass velocity at the outface flow.
It should be mentioned here that the interest in self-similar solutions is significant due to the fact that partial differential equations can be reduced to ordinary ones.
As it was mentioned in \cite{Pascal_1988}, it is always worth searching for self-similar solutions before solving the problem numerically.

Nonlinear effects associated with unsteady flows through a porous medium are analysed in paper \cite{Pascal_1991}.
For unsteady flows of polytropic gas and slightly compressible fluid of power law behavior the author derived similar nonlinear parabolic equations for determining pressure distribution.
The exact self-similar solutions in a closed form for the first boundary value problem were presented and discussed.
The existence of a moving pressure front was established.
It has turned out that pressure disturbances in polytropic gas (or slightly compressible fluid of power law behavior) flowing through a porous medium propagate with final velocity, which is a monotonically decreasing function of time.

In paper \cite{Pascal_1997}, the authors discuss two classes of non-Darcian flows through porous media, namely the turbulent flow of polytropic gases and the flow of non-Newtonian power law fluids.
The momentum equation, governing the gas flow, is the Darcy--Forcheimer equation, while the power law fluid flow model is based on a modified Darcy's law taking into account the non-linear rheological effect on the flow behavior. In both cases the governing equations belong to a class of non-linear degenerate parabolic equations.
The solutions of these equations have characteristics of a travel wave.
In case, where the boundary conditions are power law functions of time, the authors have managed to found self-similar solutions for the problem.

Another no less important case of deviation from Darcy's law is the situation, where the linear correlation between the pressure gradient and velocity is no longer valid due to the fact, that viscosity or permeability are no longer constants.
That is discussed further in this paper.
The fact that viscosity depends on pressure has been reported in many publications.
We would like to make reference only to some of them, namely the early work by \cite{Bridgman} and some recent experiments in \cite{Bair2007, Bair2008}.
It should be noted that in general one cannot assume that permeability is independent of pressure.

For a gas flow through tight gas reservoirs the average free path of the gas molecules can no longer be neglected as compared to the average effective radius of rock pore throat.
It makes possible for gas molecules to slip along pore surfaces. Such gas slippage phenomenon creates an additional flux mechanism besides viscous flow.
This effect yields an overestimated value of permeability of the gas being measured in comparison with the true absolute permeability of a liquid.

The first studies of gas slippage in porous media were conducted by Adzumi \cite{Adzumi_1937} and later by Klinkenberg \cite{klinkenberg_1942}, who proved that there exists relationship between the measured gas permeability $k$ and the mean core pressure $p$:
\begin{equation*}
k = k_\infty \left( 1+d/p \right),
\end{equation*}

Klinkenberg observed that the gas permeability approaches a limiting value at an infinite pressure $p$.
This limiting permeability $k_\infty$ is sometimes referred to as the equivalent liquid permeability, which is also called Klinkenberg-corrected permeability.
The parameter $d$ designates the gas slippage factor, which is a constant, related to the mean free path of the gas molecules at the pressure $p$ and effective pore radius.

The subsequent works have focused on correlating parameters of the Klinkenberg gas slippage factor $d$ and the equivalent liquid permeability $k_\infty$.
Basing on the results of their core sample experiments, Heid et al.\cite{heid_1950} and Jones and Owens \cite{jones_1979} proposed similar approach in the form of correlations between $d$ and $k_\infty$ : $d = C_1 k_\infty^{C_2}$ (where the constants are the following: $C_1 = 11.419$, $C_2 = -0.39$ in [3] and $C_1 = 12.639$, $C_2 = -0.33$ in \cite{jones_1979}).

Later Sampath and Keighin \cite{sampath_1981} studied core samples from a tight gas sand field and proposed a formula relating the gas slippage factor $d$ to the ratio of Klinkenberg-corrected permeability $k_\infty$ to effective porosity $\phi$: $d = C_1 \left( k_\infty/\phi \right)^{C_2}$ (corresponding constants are the following: $C_1 = 13.851$, $C_2 = -0.53$).
This correlation is interesting since theoretical analysis, reported by \cite{civan_2010} and \cite{florence_2007}, establishes "square-root" correlation
$\left( k_\infty/\phi \right)^{-0.5}$, which is close to $\left( k_\infty/\phi \right)^{C_2}$ , where $C_2 = -0.53$.
As time passed,numerous attempts were made to investigate gas flow in tight porous media.

For example, Javadpour et al. \cite{javadpour_2009} found that the behavior of the gas flow in shales deviates from the behavior described by the Fick's and Darcy's laws.
Publication \cite{rushing_2007} presents results of a laboratory research into the effects of different velocity and pressure testing conditions on a steady-state flow measurements in tight gas sands.
The authors also compared measurements of Klinkenberg-corrected permeability using both conventional steady-state technique and unsteady-state permeameters.
In the publication \cite{tanikawa_2006}, intrinsic permeability of sedimentary rocks was measured by using the nitrogen gas and water as pore fluids.

The permeability to the nitrogen gas was 2 to 10 times larger than to water on the same specimen.
The permeability to the nitrogen gas decreases with an increase of pore pressure, and the correlation between permeability to the gas and pressure could be described by Klinkenberg's formula for more experimental data.

It has been shown that we need to apply pore pressure above 1MPa, if we want to avoid Klinkenberg effect.
Tight gas and shale gas reservoir systems pose a tremendous potential resource for future development.

That is why various attempts have been made to model such kind of flow behavior.
Skjetne and Gudmundsson \cite{skjetne_1995} and Skjetne and Auriault \cite{skjetne_1999} theoretically investigated the wall-slip gas flow phenomenon in porous media based on the Navier-Stokes equation, but did not offer any correlation for the Klinkenberg effect.
Florence et al. \cite{florence_2007} made an attempt to derive a general expression for gas permeability of tight porous media, using the Hagen-Poiseuille-type equation.
Civan \cite{civan_2010} pointed out that by rigorous application of a unified Hagen-Poiseuille-type equation he made some critical improvements, which allowed him to get an accurate and effective correlation of apparent gas permeability in tight porous media.
Freeman et al. \cite{freeman_2011} examined the effects of Knudsen diffusion on gas composition in ultra-tight rock.

In the publication \cite{wu_1998}, the authors provided the analysis of steady state and transient gas flow through porous media. They have derived a convenient form of a nonlinear governing equation, incorporating the Klinkenberg effect. Under conditions of a steady state flow the governing equation becomes linear and can be solved analytically. In the case of a transient flow the authors perform the procedure of linearization, in view to find analytical solutions.

In this work we study gas flow through tight porous media. The deviation from linear correlation between the pressure gradient and velocity appears due to the fact, that permeability and viscosity are considered to be functions of pressure.
Nonlinear governing equation, incorporating Klinkenberg effect, is solved in for a transient gas flow. As a result of our analysis, approximate self-similar solutions have been found for the appropriate boundary-value problems.

An applied method can be used for finding exact solutions of integrable equations. It can be particularly effective in finding solutions for the equations for which the Cauchy problem is not solved in general case (see, for a example, books \cite{Zeldovich, Kudryashov_book}). The same approach can be successfully used for modeling of the processes of a nonlinear heat transfer  due to the similarity of mathematical description of the processes of gas filtration and those of heat transfer, as it is done in \cite{Kudr2005, Kudr2007}.

\section{The boundary value problem for the gas flow in a porous medium}

Under the isothermal conditions a gas flow in porous media is governed by a mass balance equation
\[\phi\frac{\partial\rho}{\partial t}+\nabla(\rho v)=0,\]
where $\rho$ is the gas density, $\phi$ is the porosity, $v$ is the Darcy's velocity of the gas phase defined as
\[v=-\frac{\beta}{\eta} \nabla {p}.\]
In the last relationship $\eta$ stands for viscosity, $\beta$ for permeability and $\nabla {p}$ for the pressure gradient (the gravity effects are ignored). The correlation between the gas density and the pressure is $\rho=\frac{Mp}{RT}$, where $M$ is the molar mass of the gas, $T$ is the temperature and $R$ is the universal gas constant. The combination of these equations will give the gas flow equation in the following form $\frac{\partial p}{\partial t}=\frac{1}{\phi \eta} \nabla(p\beta \nabla p)$.
Let us consider that we consider one dimentional case of a gaseous flow in a porous medium. It can be described by the following nonlinear partial differential equation
\begin{equation}
\label{I_1}
    \frac{\partial u}{\partial t}=\lambda\, \frac{\partial }{\partial x}\left(u\,K(u)\frac{\partial u}{\partial x}\right), \quad
    u(x,t) \ge 0,
\end{equation}
where $x$ is the coordinate, $t$ is the time,  $u$ is the pressure $p$, $K=\frac{\beta}{\eta}$ is the permeability of the medium, $\lambda=\frac{1}{\phi}$ is a coefficient.
Let us suppose that
\begin{equation}
\label{I_4}
    u(x\rightarrow \infty,t)=0.
\end{equation}
Assuming that at $x=0$ we get the constant pressure $u_{1}$ as
\begin{equation}
\label{I_2}
    u(x=0,t)=u_1,
\end{equation}
and the initial value for pressure as
\begin{equation}
\label{I_3}
    u(x,t=0)=0.
\end{equation}
we get the boundary-value problem for Eq.\eqref{I_1} in $x\in[0,\infty)$ .

\section{The procedure for constructing self-similar solutions for the first boundary value problem}

Using  the  change of variables
\begin{equation}
\label{M_3}
    u=A\,u^{'}, \qquad
    x=L\,x^{'}, \qquad
    t=T\,t^{'},
\end{equation}
where $A$, $L$ and $T$ are constants, from Eq.\eqref{I_1} we obtain the equation in the following form (the primes are omitted)
\begin{equation}
\label{M_4}
    \frac{\partial u}{\partial t}=A\frac{T\,\lambda\,}{L^2} \frac{\partial }{\partial x}\left(u\,K(A\,u)\frac{\partial u}{\partial x}\right).
\end{equation}

Assuming that
\begin{equation}
\label{M_5}
    A=1, \quad
    \frac{T}{L^2}=1,
\end{equation}
we can see that Eq.\eqref{I_1} admits the following transformation group
\begin{equation}
\label{M_5a}
    u=u^{'}, \qquad x=x^{'}\,e^{2a},\qquad t=t^{'}\,e^{a}.
   \end{equation}

Therefore, Eq.\eqref{I_1} is invariant under  the dilatation group of transformations with the invariants in the form
\begin{equation}
\label{M_6}
    I_1=u,\quad
    I_2=\frac{x}{\sqrt{t}}.
\end{equation}

Taking the invariants \eqref{M_6} into account we can search for exact solutions of Eq.\eqref{I_1} in the form
\begin{equation}
\label{M_7}
    u(x,t)=u_1\,f(z),\quad z=\frac{x}{\sqrt{2\lambda t}}.
\end{equation}

In this case, we have the equation for $f(z)$ in the form of nonlinear ordinary differential equation
\begin{equation}
\label{M_8}
    \frac{d}{d z}\left(f\,k(f)\,\frac{d f}{d z}\right)+z\,\frac{d f}{d z}=0, \quad
    f(z) \ge 0,
\end{equation}
where
\begin{equation*}
    k(f) = u_1 K(u_1 f),
\end{equation*}
and boundary conditions appears as
\begin{equation}
\label{M_8a}
    f(z=0)=1,\quad f(z\rightarrow\infty)=0.
\end{equation}
Eq.\eqref{M_8} does not integrable analytically.

\section{Approximate solutions for nonlinear boundary value problems}

As it was already discussed in \cite{Zeldovich, Kudryashov_book} the propagation velocity of a wave front for the pressure is finite in a number of problems of nonlinear gas filtration. This assumption corresponds to the fact that there exists a coordinate of a wave front $z=\alpha$ such that
\begin{equation}
\label{S_2}
    f(z=\alpha)=0, \quad
    \left. \frac{df}{dz} \right|_{z=a} \neq 0.
\end{equation}

Let us search for the approximate solution of Eq.\eqref{M_8} in the form
\begin{equation}
\label{S_1}
    f(z)=
    \begin{cases}
        \sum_{m=1}^{N}\,F_m\,(z-\alpha)^m, & z < \alpha, \\
        0, & z \ge \alpha.
    \end{cases}
\end{equation}

After differentiation we obtain From Eq.\eqref{S_1}
\begin{equation*}
\label{S_3}
    \left. \frac{d^m f}{d z^m} \right|_{z=\alpha} = m!\,F_m.
\end{equation*}

On the one hand, we get
\begin{equation*}
\label{S_3}
    F_m=\frac{1}{m!} \left. \frac{d^m f}{d z^m} \right|_{z=\alpha}.
\end{equation*}

But, on the other hand, from Eq.\eqref{M_8} we obtain
\begin{equation}
\label{S_5}
    F_1 = \left. \frac{df}{dz} \right|_{z=\alpha} = - \frac{\alpha}{k(0)} = - \frac{\alpha}{k_0}.
\end{equation}

Consequently, by differentiating \eqref{M_8} with respect to $z$, we find higher order derivatives of $f(z)$ at $z=\alpha$ and thus obtain the other values of coefficients $F_{m}$:
\begin{equation}
\label{S_6}
\begin{aligned}
    &F_2 = - \frac{1}{2} \left(
        \frac{1}{2\,k_0}
        + \frac {\alpha^2 k_1}{k_0^3}
    \right), \\
    &F_3 = - \frac{1}{6} \left(
        \frac{1}{12\,\alpha\,k_0}
        + \frac{5\,\alpha\,k_1}{3\,k_0^3}
        + \frac{3\,\alpha^3 k_1^{2}}{k_0^5}
        - \frac{\alpha^3 k_2}{k_0^4}
    \right), \\
    &F_4 = \frac{1}{24} \left(
        \frac {1}{24\,\alpha^2 k_0}
        - \frac{5\,k_1}{4\,k_0^3}
        - \frac{61\,\alpha^2 k_1^2}{6\,k_0^5}
        + \frac {13\,\alpha^{2} k_2}{4\,k_0^4}
        - \frac{15\,\alpha^4 k_1^3}{k_0^7}
        + \frac{10 \alpha^4 k_1\,k_2}{k_0^6}
        - \frac {\alpha^4 k_3}{k_0^5}
    \right), \\
    &\begin{split}
    F_5 = & - \frac{1}{120} \left(
        \frac{11}{720 \alpha^3 k_0}
        + \frac{7 k_1}{36 \alpha k_0^3}
        + \frac{746 \alpha k_1^2}{45 k_0^5}
        - \frac{617 \alpha k_2}{120 k_0^4}
        + \frac{85  \alpha^3 k_1^3}{k_0^7}
        - \frac{661 \alpha^3 k_1 k_2}{12 k_0^6}
    \right. \\
    &\left.
        + \frac{53 \alpha^3 k_3}{10 k_0^5}
        + \frac{105 \alpha^5 k_1^4}{k_0^9}
        - \frac{105 \alpha^5 k_1^2 k_2}{k_0^8}
        + \frac{15 \alpha^5 k_1 k_3}{k_0^7}
        + \frac{10 \alpha^5 k_2}{k_0^7}
        - \frac{\alpha^5 k_4}{k_0^6}
    \right),
    \end{split}
\end{aligned}
\end{equation}
and so on. Note, that in expressions \eqref{S_5}, \eqref{S_6} we have introduced the following notations
\begin{equation}
\label{S_9}
    k_0 = k(f(z=\alpha))=k(0),\quad
    k_1 = \left. \frac{d k}{d f} \right|_{f=0},\quad
    k_2 = \left. \frac {d^2 k}{d f^2} \right|_{f=0},\quad
    k_3 = \left. \frac{d^3 k}{df^3} \right|_{f=0}.
\end{equation}

This way we obtain an approximate solution of Eq.\eqref{M_8} for $z<\alpha$ in the form
\begin{multline}
\label{S_10}
    f(z) = \frac{\alpha}{k_0}\,(\alpha-z)
        - \frac{1}{2} \left( \frac{1}{2\,k_0} + \frac{\alpha^2 k_1}{k_0^3} \right) (\alpha-z)^2 + \\
        + \frac{1}{6} \left( \frac{1}{12\,\alpha\,k_0} + \frac{5\,\alpha\,k_1}{3\,k_0^3}
        + \frac {3\,\alpha^3 k_1^2}{k_0^5} - \frac{\alpha^3 k_2}{k_0^4} \right) (\alpha-z)^3 + \\
        + \frac{1}{24} \left( \frac{1}{24\,\alpha^2 k_0} - \frac{5\,k_1}{4\,k_0^3}
        - \frac{61\,\alpha^2 k_1^2}{6\,k_0^5} + \frac{13\,\alpha^2 k_2}{4\,k_0^4}
        + \frac{10 \alpha^4 k_2\,k_1}{k_0^6} - \right. \\
    - \left. \frac{15\,\alpha^4 k_1^3}{k_0^7} - \frac{\alpha^4 k_3}{k_0^5} \right) (\alpha-z)^4
    + \frac{1}{120} \left(
        \frac{11}{720 \alpha^3 k_0}
        + \frac{7 k_1}{36 \alpha k_0^3}
        + \frac{746 \alpha k_1^2}{45 k_0^5}
        - \frac{617 \alpha k_2}{120 k_0^4} +
    \right. \\
        + \frac{85  \alpha^3 k_1^3}{k_0^7}
        - \frac{661 \alpha^3 k_1 k_2}{12 k_0^6}
        + \frac{53 \alpha^3 k_3}{10 k_0^5}
        + \frac{105 \alpha^5 k_1^4}{k_0^9}
        - \frac{105 \alpha^5 k_1^2 k_2}{k_0^8}
        + \frac{15 \alpha^5 k_1 k_3}{k_0^7} + \\
    \left.
        + \frac{10 \alpha^5 k_2}{k_0^7}
        - \frac{\alpha^5 k_4}{k_0^6}
    \right) (\alpha-z)^5
    + \dotsb,
\end{multline}
and so on.

The value of the constant $\alpha$ may be obtained from the first one of boundary conditions \eqref{M_8a}.
By substituting the approximate solution \eqref{S_10} in this condition we get (in case of $k(f)=K(u_1 f) \ne const$)
\begin{gather}
\label{S_11}
    1 = f(0) \sim \alpha^2 + \alpha^4 + \alpha^6 + \dotsb .
\end{gather}
By solving this nonlinear equation numericallywe obtain the value of $\alpha$ and according to\eqref{S_1}, \eqref{S_6} we obtain the approximate solution of the Eq.\eqref{I_1}.

\section{Examples of approximate and numerical solutions for different cases of $k(f)$}

Let us consider examples of equations with different values of $k(f)$ i.e. with different cases of dependence of the permeability and viscosity on the pressure. These cases should be regarded as various possible approximations of real dependences, resulting from experiments. The corresponding parameters in the following formulas have been chosen to be fitting parameters.  In these examples we are looking for approximate self-similar solutions of the problem \eqref{I_1} with appropriate boundary conditions, using \eqref{M_7},\eqref{S_9},\eqref{S_10}.
\begin{gather}
\label{E_0}
\frac{d}{d z} \left( f\,k(f)\,\frac{df}{dz} \right) + z\,\frac{df}{dz}=0, \\
\\
f(z=0)=1, \qquad
f(z=\alpha)=0, \qquad f'(z=\alpha)=-\alpha.
\end{gather}
Our approximate self-similar solution is found by considering only five approximation coefficients (see \eqref{S_1}, \eqref{S_6}), but our approach allows us to look for much more than that.

In order to check our approximate self-similar solutions we also used the numerical approach. We applied the finite differences, shooting method, Newton's method and linear extrapolation at a small $f(z)$ to obtain the numerical solution. In our calculations the step size was $10^{-4}$. Calculating by shooting method was proceed until $f'(\alpha)+\alpha < 10^{-3}$. The comparison of the numerical and approximate analytical results is presented in Figures 1--5.

\emph{Example 1.}
Let $K(u)$ be constant. Without the loss of generality let us assume that
\begin{equation*}
    K(u)=\frac{1}{u_1}.
\end{equation*}
In this case we can rewrite Eq.\eqref{I_1} in the form
\begin{equation}
\label{E_1}
    \frac{\partial u}{\partial t}=\frac{\lambda}{u_1}\, \frac{\partial }{\partial x}\left(u\,\frac{\partial u}{\partial x}\right),
\end{equation}

The value $k_n$ in \eqref{S_9} becomes
\begin{equation}
\label{E_2}
    k_0 = 1,\qquad
    k_n = 0 \quad (n \ge 1).
\end{equation}
By taking into account\eqref{M_7}, \eqref{S_10}, \eqref{E_2}, the approximate self-similar solution of the boundary-value problem for the Eq.\eqref{E_1} takes the form \cite{Kudr2005, Kudr2007}
\begin{gather*}
    u(x,t) = u_1 f \left( \frac{x}{\sqrt{2 \lambda t}} \right), \\
    \label{E_2a}
    f(z) = \alpha \, (\alpha-z)
        - \frac{1}{4} \,(\alpha-z)^2 + \frac{1}{72\,\alpha} \,(\alpha-z)^3
        + \frac{1}{576\,\alpha^2}\,(\alpha-z)^4 + \frac{11}{86400 \alpha^3} (\alpha-z)^5 + \dotsb .
\end{gather*}
From condition \eqref{S_2} we have $\alpha = 1.143$.

\begin{figure}
\psfrag{xlbl}{$z$}
\psfrag{ylbl}{$f(z)$}
\center{\includegraphics[width=0.7\linewidth]{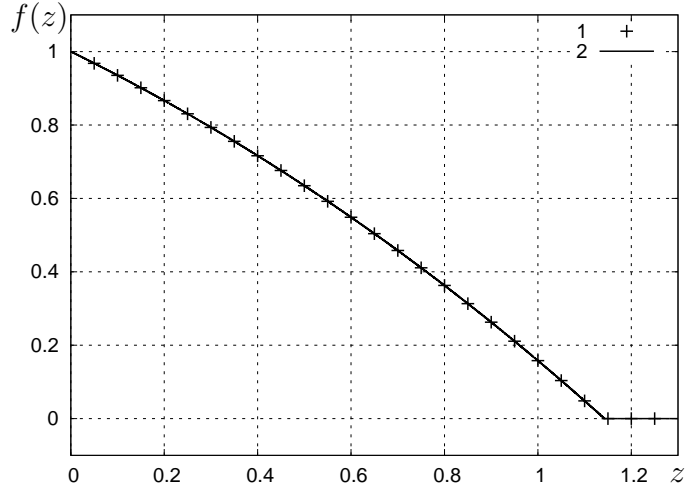}}
\caption{Solution for the problem \eqref{M_8}, \eqref{M_8a} for \emph{Example 1} (line 1 --- numerical solution, line 2 --- exact solution \eqref{E_2a})}
\label{fig:const}
\end{figure}

\emph{Example 2.}
Let us consider the linear case of the dependence $K(u)$.
Without the loss of generality let us present it in the form
\begin{equation}
\label{E_3}
    K(u) = \frac{1}{u_1} \left( 1 +\frac{b}{u_1} u \right),
\end{equation}
where $b \ne 0$ is some constant.
Then the Eq.\eqref{I_1} takes the form
\begin{equation}
\label{E_4}
    \frac{\partial u}{\partial t} = \frac{\lambda}{u_1}\,\frac{\partial}{\partial x}\left(
        u\,\left( 1 + \frac{b}{u_1}\,u \right)\,\frac{\partial u}{\partial x}
    \right),
\end{equation}
and the values of $k_n$ in \eqref{S_9} become
\begin{equation}
\label{E_5}
    k_0 = 1,\qquad
    k_1 = b,\qquad
    k_n = 0 \quad (n \ge 2).
\end{equation}

Taking \eqref{M_7},\eqref{S_10}, \eqref{E_5} into account we get approximate self-similar solution for the Eq.\eqref{E_4}
\begin{gather*}
    u(x,t) = u_1 f \left( \frac{x}{\sqrt{2 \lambda t}} \right), \\
\label{E_5a}
\begin{split}
    f(z) = \; & \alpha (\alpha-z)
        - \frac{1}{4} \left( 2 \alpha^2 b + 1 \right) (\alpha-z)^2 + \\
        &+ \frac{1}{72 \alpha} \left( 2 \alpha^2 b + 1 \right) \left( 18 \alpha^2 b + 1 \right) (\alpha-z)^3 + \\
        &+ \frac{1}{576 \alpha^2} \left( 2 \alpha^2 b + 1 \right) \left( 180 \alpha^4 b^2 + 32 \alpha^2 b - 1 \right) (\alpha-z)^4 + \\
        &+ \frac{1}{86400 \alpha^3} \left( 2 \alpha^2 b + 1 \right)
            \left( 37800 \alpha^6 b^3 + 11700 \alpha^4 b^2 + 118 \alpha^2 b + 11 \right) (\alpha-z)^5 + \dotsb .
\end{split}
\end{gather*}
Let us find the value of $\alpha$ from the boundary condition \eqref{S_2}. The results of numerical solution for various values of parameter $b$ are presented in Table \ref{tabl:example2}.
\begin{table}
    \renewcommand{\arraystretch}{1.3}
    \caption{Values of parameter $\alpha$ for various values of $b$ in \eqref{E_4}}
    \label{tabl:example2}
\begin{center}
\begin{tabular}{|c|c|c|c|c|c|c|c|c|}
    \hline
        $b$         &$-0.2$     &$-0.15$    &$-0.1$     &$-0.05$    &$0.05$     &$0.1$      &$0.15$     \\
    \hline
        $\alpha$    &$1.088$    &$1.102$    &$1.116$    &$1.129$    &$1.156$    &$1.169$    &$1.182$    \\
    \hline
\end{tabular}
\end{center}
\end{table}

\begin{figure}
\psfrag{xlbl}{$z$}
\psfrag{ylbl}{$f(z)$}
\center{\includegraphics[width=0.7\linewidth]{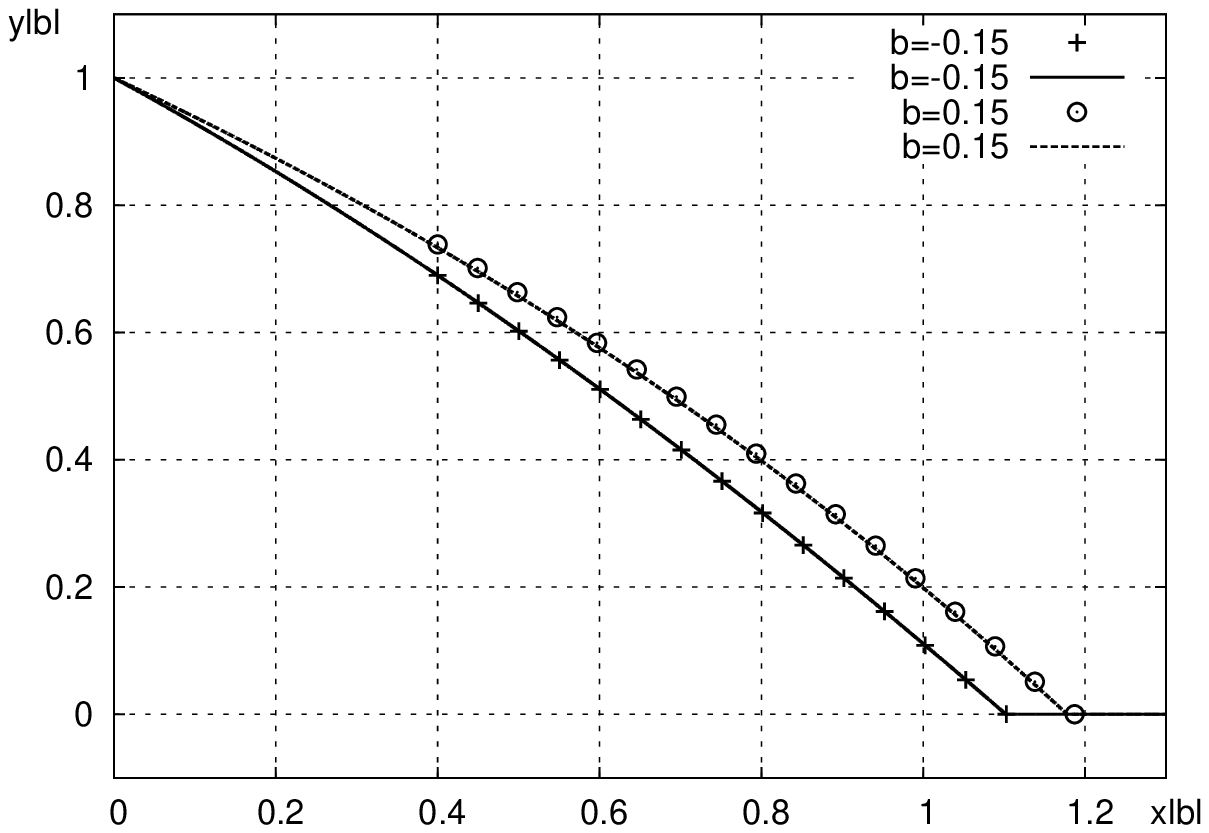}}
\caption{Solution for the problem \eqref{M_8}, \eqref{M_8a} for \emph{Example 2} for various values of $b$ (points --- numerical solution, lines --- exact solution \eqref{E_5a})}
\label{fig:linear}
\end{figure}

Note that in case of $b=-\frac{1}{2\alpha^2}$ we have an exact solution of Eq.\eqref{E_4} in the form
\begin{gather*}
    u(x,t) = u_1 f \left( \frac{x}{\sqrt{2 \lambda t}} \right), \\
    f(z) = \alpha (\alpha-z).
\end{gather*}
From the condition at $z=0$ (see \eqref{M_8a}) we have $\alpha=1,$ and therefore $b = - \frac{1}{2}$.
The coordinate of the wave front in this case is
\begin{equation*}
    x = \sqrt{2 \lambda t}.
\end{equation*}

\emph{Example 3.}
Let us suppose that
\begin{equation}
\label{E_6}
    K(u) = \frac{1}{u_1} \left( \exp{\left(\frac{b}{u_1}\,u \right)} \right),\qquad \qquad
\end{equation}
where $b \ne 0 $ is some constant.
Then we have Eq.\eqref{I_1} in the form
\begin{equation}
\label{E_7}
    \frac{\partial u}{\partial t} = \frac{\lambda}{u_1}\,\frac{\partial}{\partial x}\left(
        u\,\exp{\left( \frac{b}{u_1}\,u \right)}\, \frac{\partial u}{\partial x}
    \right),\qquad
\end{equation}

The values of $k_n$ in \eqref{S_9} become
\begin{equation*}
\label{E_8}
    k_n = b^n \quad (n \ge 0).
\end{equation*}
Then we have an approximate self-similar solution of the boundary-value problem for the Eq.\eqref{E_7} in the form
\begin{gather*}
    u(x,t) = u_1 f \left( \frac{x}{\sqrt{2 \lambda t}} \right), \\
\label{E_8a}
\begin{split}
    f(z) = \; & \alpha (\alpha-z)
        - \frac{1}{4} \left( 2 \alpha^2 b + 1 \right) (\alpha-z)^2 + \\
        &+ \frac{1}{72 \alpha} \left( 24 \alpha^4 b^2 + 20 \alpha^2 b + 1 \right) (\alpha-z)^3 + \\
        &- \frac{1}{576 \alpha^2} \left( 144 \alpha^6 b^3 + 166 \alpha^4 b^2 + 30 \alpha^2 b - 1 \right) (\alpha-z)^4 + \\
        &+ \frac{1}{86400 \alpha^3} \left( 17280 \alpha^8 b^4 + 25356 \alpha^6 b^3 + 8234 \alpha^4 b^2 + 140 \alpha^2 b + 11 \right) (\alpha-z)^5 + \dotsb.
\end{split}
\end{gather*}
From the boundary condition \eqref{M_8a} we obtain nonlinear equation for $\alpha$.
The results of numeric solution of this equation for various values of parameter $b$ are shown in Table \ref{tabl:example3}.
\begin{table}
    \renewcommand{\arraystretch}{1.3}
    \caption{Values of parameter $\alpha$ for various values of $b$ in \eqref{E_7}}
    \label{tabl:example3}
\begin{center}
\begin{tabular}{|c|c|c|c|c|c|c|c|c|}
    \hline
        $b$         &$-0.3$     &$-0.25$    &$-0.2$     &$-0.15$    &$-0.1$     &$-0.05$    &$0.05$     &$0.1$      \\
    \hline
        $\alpha$    &$1.068$    &$1.080$    &$1.092$    &$1.104$    &$1.117$    &$1.130$    &$1.156$    &$1.169$    \\
    \hline
\end{tabular}
\end{center}
\end{table}

\begin{figure}
\psfrag{xlbl}{$z$}
\psfrag{ylbl}{$f(z)$}
\center{\includegraphics[width=0.7\linewidth]{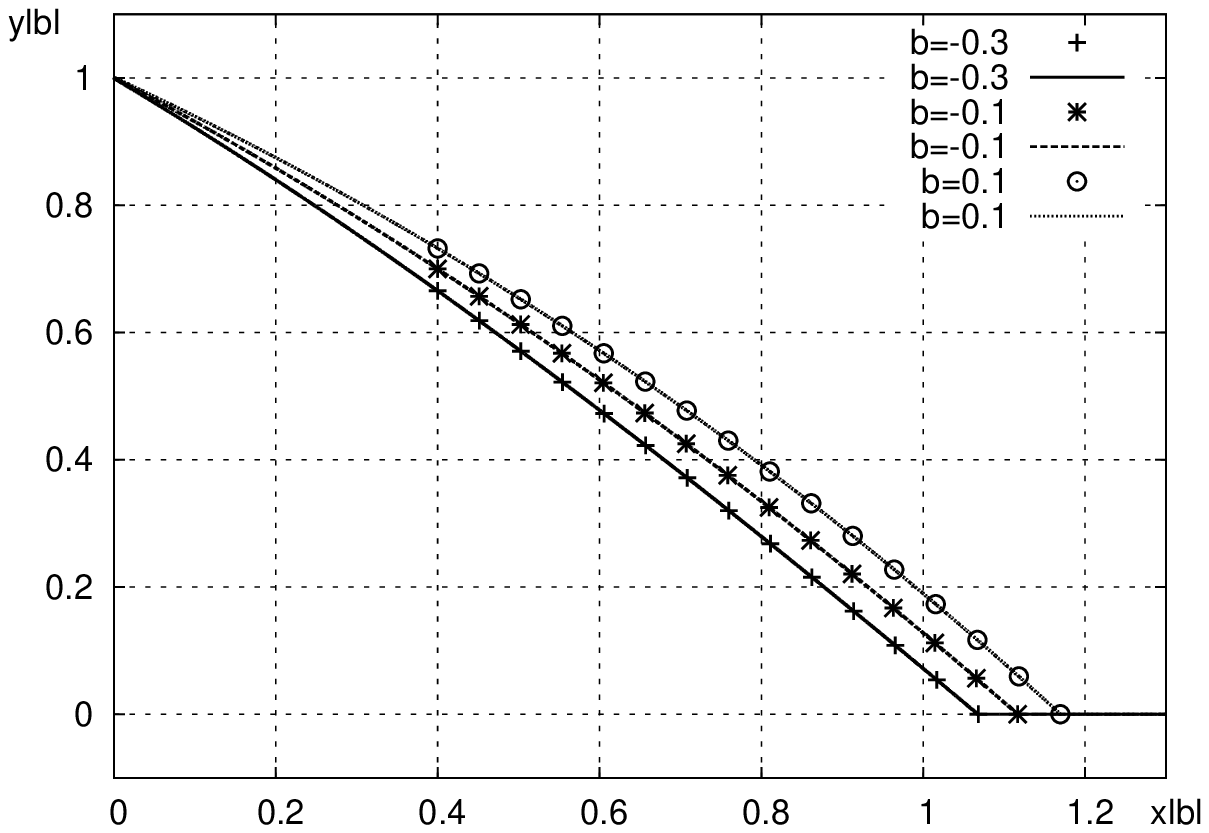}}
\caption{Solution for the problem \eqref{M_8}, \eqref{M_8a} for \emph{Example 3} for various values of $b$ (points --- numerical solution, lines --- exact solution \eqref{E_8a})}
\label{fig:exp}
\end{figure}

\emph{Example 4.}
Let us suppose that
\begin{equation}
\label{E_9}
    K(u) = \frac{1}{u_1} \left( 1 + \frac{b}{u_1}\, u \right)^{-1},\qquad \qquad
\end{equation}
where $b \ne 0$ is some constant.
Then we have Eq.\eqref{I_1} in the form
\begin{equation}
\label{E_10}
    \frac{\partial u}{\partial t} = \frac{\lambda}{u_1}\,\frac{\partial}{\partial x}\left(
     \frac{  u}{1 + m_1 u}\,\frac{\partial u}{\partial x}
    \right), \qquad m_1=\frac{b}{u_1},
\end{equation}

For values of $k_n$ in \eqref{S_9} we have
\begin{equation*}
\label{E_11}
    k_n = (-1)^n \, n! \, b^n \quad (n \ge 0).
\end{equation*}
Thus we obtain an approximate self-similar solution of the boundary-value problem for the Eq.\eqref{E_10} in the form
\begin{gather}
\nonumber
    u(x,t) = u_1 f \left( \frac{x}{\sqrt{2 \lambda t}} \right), \\
\label{E_12}
\begin{split}
    f(z) = \; & \alpha (\alpha-z)
        - \frac{1}{4} \left( 1 - 2 \alpha^2 b \right) (\alpha-z)^2 + \\
        &+ \frac{1}{72 \alpha} \left( 12 \alpha^4 b^2 - 20 \alpha^2 b + 1 \right) (\alpha-z)^3 + \\
        &+ \frac{1}{576 \alpha^2} \left( 24 \alpha^6 b^3 - 88 \alpha^4 b^2 + 30 \alpha^2 b + 1 \right) (\alpha-z)^4 + \\
        &+ \frac{1}{86400 \alpha^3} \left( 720 \alpha^8 b^4 - 4776 \alpha^6 b^3 + 4532 \alpha^4 b^2 - 140 \alpha^2 b + 11 \right) (\alpha-z)^5 + \dotsb.
\end{split}
\end{gather}

Substituting the approximate solution \eqref{E_12} into the boundary condition \eqref{M_8a} we obtain nonlinear equation for $\alpha$ determination. The numeric solutions for this equation for various values of parameter $b$ are shown in Table \ref{tabl:example4}.
\begin{table}
    \renewcommand{\arraystretch}{1.3}
    \caption{Values of parameter $\alpha$ for various values of $b$ in \eqref{E_10}}
    \label{tabl:example4}
\begin{center}
\begin{tabular}{|c|c|c|c|c|c|c|c|c|c|}
    \hline
        $b$         &$-0.15$    &$-0.1$     &$-0.05$    &$0.05$     &$0.1$      &$0.15$     &$0.2$      &$0.25$     &$0.3$      \\
    \hline
        $\alpha$    &$1.186$    &$1.171$    &$1.156$    &$1.130$    &$1.117$    &$1.106$    &$1.095$    &$1.084$    &$1.074$    \\
    \hline
\end{tabular}
\end{center}
\end{table}

\begin{figure}
\psfrag{xlbl}{$z$}
\psfrag{ylbl}{$f(z)$}
\center{\includegraphics[width=0.7\linewidth]{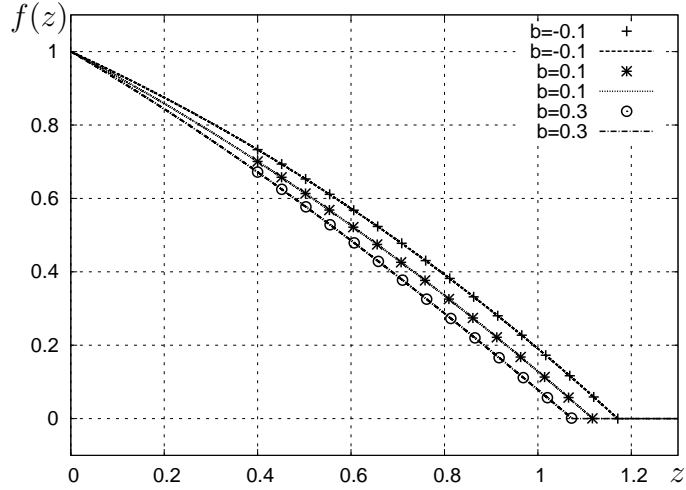}}
\caption{Solution for the problem \eqref{M_8}, \eqref{M_8a} for \emph{Example 4} for various values of $b$ (points --- numerical solution, lines --- exact solution \eqref{E_12})}
\label{fig:inverse_1}
\end{figure}

\emph{Example 5.}
Let us consider the following case of permeability
\begin{equation}
\label{E_13}
    K(u) = \frac{1}{u_1} \frac{1+m_1 u}{1+m_2 u},\qquad m_1=\frac{b}{u_1}\qquad m_2=\frac{c}{u_1} \qquad
    b c \ne 0, \qquad
    b \ne c,
\end{equation}
where $b$, $c$ are some constants.
In case of $b=0$ or $c=0$ we get \emph{Example 4} and \emph{Example 2} accordingly.
If $b=c$ then $K(u)$ becomes $1$ (\emph{Example 1}).
Then we get Eq.\eqref{I_1} in the form
\begin{equation}
\label{E_14}
    \frac{\partial u}{\partial t} = \frac{\lambda}{u_1}\,\frac{\partial}{\partial x} \left(
      u\, \left( \frac{1+m_1 u}{1+m_2 u}\right) \frac{\partial u}{\partial x}
    \right), \qquad m_1=\frac{b}{u_1}\qquad m_2=\frac{c}{u_1}
\end{equation}

For values of $k_n$ \eqref{S_9} we obtain
\begin{equation*}
\label{E_15}
    k_0 = a-b, \qquad
    k_n = (-1)^{n-1} \, n! \, b^{n-1} (a-b) \quad (n \ge 1).
\end{equation*}
Then we get an approximate self-similar solution for the boundary-value problem for the Eq.\eqref{E_14} in the form
\begin{gather}
\nonumber
    u(x,t) = u_1 f \left( \frac{x}{\sqrt{2 \lambda t}} \right), \\
\label{E_16}
\begin{split}
    f(z) = \; & \alpha (\alpha-z)
        - \frac{1}{4} \left( 1 + 2 \alpha^2 (b-c) \right) (\alpha-z)^2 + \\
        &+ \frac{1}{72 \alpha} \left( 36 \alpha^4 b^2 - 48 \alpha^4 b c + 12 \alpha^4 c^2 + 20 \alpha^2 (b-c) + 1 \right) (\alpha-z)^3 + \\
        &+ \frac{1}{576 \alpha^2} \left(
        - 360 \alpha^6 b^3 + 600 \alpha^6 b^2 c - 264 \alpha^6 b c^2 + 24 \alpha^6 c^3 - 244 \alpha^4 b^2 + 332 \alpha^4 b c - \right. \\
        &\left. - 88 \alpha^4 c^2 - 30 \alpha^2 (b-c) + 1 \right) (\alpha-z)^4
        + \frac{1}{86400 \alpha^3} \left( 75600 \alpha^8 b^4 - 151200 \alpha^8 b^3 c + \right. \\
        &+ 93600 \alpha^8 b^2 c^2 - 18720 \alpha^8 b c^3 + 720 \alpha^8 c^4 + 61200 \alpha^6 b^3 - 104280 \alpha^6 b^2 c + \\
        & + 47856 \alpha^6 b c^2 - 4776 \alpha^6 c^3 + 11936 \alpha^4 b^2 - 16468 \alpha^4 b c + 4532 \alpha^4 c^2 + 140 \alpha^2 b - \\
        &\left. - 140 \alpha^2 c + 11 \right) (\alpha-z)^5 + \dotsb.
\end{split}
\end{gather}

From the boundary condition \eqref{M_8a} we obtain a nonlinear equation for $\alpha$.
The numeric solutions for this equation for various values of parameter $b$ and $c$ are shown in Table \ref{tabl:example5}.
\begin{table}
    \renewcommand{\arraystretch}{1.3}
    \caption{Values of parameter $\alpha$ (values, obtained from numerical solving, marked with $^*$) for various values of $b$ and $c$ in the case of pressure dependent permeability in form \eqref{E_14}}
    \label{tabl:example5}
\begin{center}
\begin{tabular}{|c|c|c|c|c|}
    \hline
        \backslashbox{$b$}{$c$}     &$-0.2$     &$-0.1$     &$0.1$              &$0.2$      \\
    \hline
        $-0.2$                      &$-$        &$1.114$    &$1.064/1.065^*$    &$1.044$    \\
    \hline
        $-0.1$                      &$1.173$    &$-$        &$1.092$            &$1.070$    \\
    \hline
        $0.1$                       &$1.217/1.231^*$        &$1.193/1.198^*$        &$-$                &$1.119$    \\
    \hline
        $0.2$                       &$-$        &$1.197/1.225^*$        &$1.165/1.167^*$    &$-$        \\
    \hline
\end{tabular}
\end{center}
\end{table}

\begin{figure}
\psfrag{xlbl}{$z$}
\psfrag{ylbl}{$f(z)$}
\center{\includegraphics[width=0.7\linewidth]{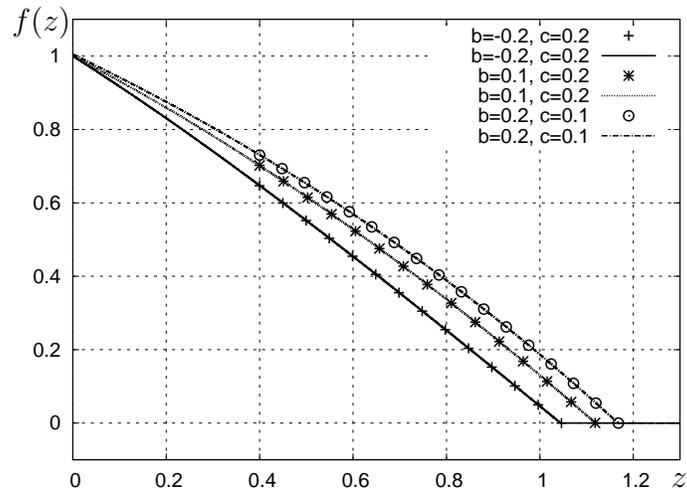}}
\caption{Solution for the problem \eqref{M_8}, \eqref{M_8a} for \emph{Example 5} for various values of $b$ and $c$ (points --- numerical solution, lines --- exact solution \eqref{E_16})}
\label{fig:inverse_2}
\end{figure}

\section{Conclusion}

Let us summarize the results of this paper. We have considered one of the boundary value problems for the gas filtration in porous medium with pressure dependent permeability. It has turned out that this problem can be solved by taking the self-similar variables into account. Thus, our problem has been transformed into the boundary value problem for a nonlinear ordinary differential equation, for which we have managed to find approximate self-similar solutions. In order to check our results we have also performed numerical integration and shown the coincidence of the approximate and numerical solutions. The value of approximate analytical solutions is that they can be used to test the results of numerical simulation, eliminate errors and improve the stability of numerical procedures. In addition, it is worth noting that approximate analytical solutions are found to be very useful by themselves. They make it easy to understand the physics of the processes and the contribution of various parameters of the problem being solved. It should be noted that our approach can also be applied in finding the solution for other nonlinear problems, particularly for problems of a nonlinear heat transfer.

\section{Acknowledgements}

This research was partially supported by Federal Target Programmes
``Research and Scientific---Pedagogical Personnel of Innovation in Russian Federation on 2009-�2013'' and ``Researches and developments in priority directions of development of a scientifically-technological complex of Russia on 2007--2013''.

\end{document}